\documentclass[11pt]{article}

\usepackage{epsfig}

\newcommand{\be}{\begin{equation}}
\newcommand{\ee}{\end{equation}}
\newcommand{\bea}{\begin{array}}
\newcommand{\ea}{\end{array}}
\newcommand{\beqa}{\begin{eqnarray}}
\newcommand{\eeqa}{\end{eqnarray}}
\newcommand{\bean}{\begin{eqnarray*}}
\newcommand{\eean}{\end{eqnarray*}}

\def\up#1{\leavevmode \raise.16ex\hbox{#1}}

\def\sqr#1#2{{\vcenter{\vbox{\hrule height.#2pt
        \hbox{\vrule width.#2pt height#1pt \kern#1pt
          \vrule width.#2pt}
        \hrule height.#2pt}}}}

\newcommand{\gapproxeq}{\lower .7ex\hbox{$\;\stackrel{\textstyle >}{\sim}\;$}}
\newcommand{\lapproxeq}{\lower .7ex\hbox{$\;\stackrel{\textstyle <}{\sim}\;$}}

\def\thebibliography#1{{\bf REFERENCES\markboth
 {REFERENCES}{REFERENCES}}\list
 {[\arabic{enumi}]}{\settowidth\labelwidth{[#1]}\leftmargin\labelwidth
 \advance\leftmargin\labelsep
 \usecounter{enumi}}
 \def\newblock{\hskip .11em plus .33em minus -.07em}
 \sloppy
 \sfcode`\.=1000\relax}

\begin{document}
\begin{flushright}
DSF 27-2002  \\ quant-ph/0301017\\
\end{flushright}
\vspace{1cm}

\begin{center}
{\bf \large{ON INTERFEROMETRIC DUALITY
IN MULTIBEAM EXPERIMENTS}}\\
\vspace{1cm}
G. Bimonte and R. Musto
\end{center}
\vspace{1cm}

\begin{center}
{\it   Dipartimento di Scienze Fisiche, Universit\`{a} di Napoli, Federico
II\\Complesso Universitario MSA, via Cintia, I-80126, Napoli, Italy;
\\ INFN, Sezione di Napoli, Napoli, ITALY.\\
\small e-mail: \tt bimonte,musto@napoli.infn.it } \\
\end{center}

\begin{abstract}

We critically analyze the problem of formulating duality between
fringe visibility and which-way information, in multibeam
interference experiments. We show that the traditional notion of
visibility is incompatible with any intuitive idea of
complementarity, but for the two-beam case.  We derive a number of
new inequalities, not present in the two-beam case, one of them
coinciding with a recently proposed  multibeam generalization of
the inequality found by Greenberger and YaSin. We show, by an
explicit procedure of optimization in a three-beam case, that
suggested generalizations of Englert's inequality, do not convey,
differently from the two-beam case, the idea of complementarity,
according to which an increase of visibility is at the cost of a
loss in path information, and viceversa.

\end{abstract}

\section{Introduction}

Interferometric duality, as complementarity between fringe
visibility and which-way information is called today, has  a long,
perhaps a surprisingly long history (for a recent review, see
\cite{berg}). It was the central issue of the famous debate
between  Einstein and Bohr, on complementarity.  Even if, already
at that time, in defending  complementarity against Einstein's
criticism, Bohr pointed out that not only the system under
observation, but also the measuring apparatus should be regarded
as a quantum object \cite{wheeler}, the discussion was essentially
semiclassical in nature.  As it was based essentially on the
position-momentum Heisenberg uncertainty principle, it  considered
only  the two extreme cases, of either a purely particle-like or a
purely wave-like behavior of the system. It was  only  in 1979
that Wootters and Zurek, \cite{zurek} gave the first full quantum
mechanical treatment of Young interference, in the presence of a
which-way detector. They  recognized  that "in Einstein's version
of the double-slit experiment, one can retain a surprisingly
strong interference pattern by not insisting on a 100\% reliable
determination of the slit through which each photon passes".

By now, a consistent and simple formulation of interferometric
duality has been achieved in the case of {\it two} interfering
beams. In the absence of a which-way detector, Greenberger and
YaSin \cite{gree}, showed that it was possible to convert the
basic quantum mechanical inequality ${\rm Tr} \rho^2 \le 1$, into
one connecting the fringe visibility to the predictability of the
path, based on unequal beam populations.  This is an
experimentally testable inequality, as it involves physically
measurable  quantities. For pure states, when the inequality is
saturated, this statement becomes a formulation of interferometric
duality; any increase in predictability is at the cost of a
decrease in visibility, and vice versa.

In the case  of an interference experiment  performed in the
presence of a which-way detector,  in order to gain information on
the  path, one needs to carry out a measurement on the detector,
after the passage of each  quanton. Since, in general, no
measurement ensures an unambiguous path reconstruction, the
determination of the  best possible measurement is a matter of
statistical decision theory, that requires an {\it a priori}
choice of an   evaluation criterion. In their pioneering work,
Wootters and Zurek, \cite{zurek}, used Shannon's definition of
information entropy \cite{shannon} in order to evaluate the
which-way information gained after the measurement. Following this
suggestion, Englert \cite{englert}, by using a different criterion
for evaluating the available information, was able to establish an
inequality, stating that the sum of  the square of the {\it
distinguishability}, that gives a quantitative estimate of the
way\footnote{In Ref. \cite{englert} the distinguishability is
expressed in terms of the optimum likelihood ${\cal L}_{opt}$ for
"guessing the way right". This optimum likelihood is one minus the
optimum average Bayes cost $\bar{C}_{opt}$}, and the {\it
visibility} squared, is bound by one. Again, the inequality is
saturated for pure states, turning into a statement of
interferometric  duality: any gain in distinguishability is paid
by a loss in visibility and vice versa.

In the present paper we discuss the issue of formulating
interferometric  duality, in the case of multibeam experiments. As
an example of the problems arising, we may refer to an experiment
\cite{mei} with four beams,  in which  the surprising result is
found that scattering of a photon by one of the beams, may  lead
to an increase of visibility, rather than to an attenuation. (For
a comment see \cite{luis2}).  To get a better understanding of
this experiment, we build an analytical three-beam example, which
shows that, differently from the two-beam case, the traditional
visibility may increase, after an interaction of the beams with
another quantum system. This points towards the need for a
different notion of visibility, and one possibility is offered in
\cite{durr}, where the visibility is defined as the properly
normalized, rms deviation of the fringes intensity from its mean
value. We briefly review D\"{u}rr's \cite{durr} derivation, for
the multibeam case, of an inequality similar to the one of
Greenberger and YaSin \cite{gree}, that relates this new notion of
visibility, to a corresponding newly defined predictability.
Again, in the case of pure state the inequality is saturated and,
then, in analogy with the two beam case, may be taken as a formal
definition of interferometric duality. However, as we will discuss
later, this is at the cost of using a definition of predictability
that has some how lost contact with the ability  of guessing the
way right. Furthermore we   show how, in the multibeam case, it is
possible to construct new inequalities, resulting, like the one of
Greenberger and YaSin, from basic quantum mechanical properties of
the density matrix. Each of them can be written in terms of
quantities that, in principle, may be measured in interference
experiments, such as higher momenta of fringes intensity. The new
inequalities then provide, exactly as the original one,
independent tests on the validity of quantum mechanics in
multibeam interference experiments. They also are saturated for
pure states, but, at least at first sight, they do not seem to
convey any simple relation with the principle of complementarity.

Then we turn to the more interesting problem of complementarity in
the presence of a which-way detector. By introducing two
alternative definitions of distinguishability, D\"{u}rr
constructed a generalization of  Englert's inequality to the
multibeam case, proposing to look at it as a formal definition of
interferometric duality. We show that, apart from the two beam
case, the new inequality holds as an equality only for the extreme
cases where either the visibility or the distinguishability
vanishes, even when the beams  and the detector are both prepared
in pure states. Then,  there may be cases in which the
distinguishability and the visibility both increase or decrease at
the same time. This is in sharp contrast with the idea of
complementarity, according to which "...the more clearly we wish
to observe the wave nature ...the more information we must give up
about... particle properties" \cite{zurek}. In a recent  paper
\cite{bimonte},  an example in which this situation occurs  was
constructed. However, we considered there an extremely simplified
model for the detector, having a two-dimensional space of states.
A realistic model requires an infinite Hilbert space of states,
and we analyze it in this paper. This is a much harder problem,
because the task of determining the path distinguishability
implies the solution of an optimization problem, that has to be
performed now in an infinite dimensional space.  We report the
full proof in this paper, not only for the sake of completeness,
but also because it provides an example in quantum decision
theory, which is a subject where few general results are known,
and few cases can be actually treated. Surprisingly, in the case
we examined, the distinguishability of the infinite-dimensional
problem coincides with the one found in \cite{bimonte}, for the
simplified model. This shows that the conclusions drawn in
\cite{bimonte} have full generality,  showing that  the notion of
interferometric duality in the multibeam case has not been yet
properly formulated.

The paper is organized as follows: in Sec. II we discuss
interferometric-duality schemes, not involving which way
detectors. In Sec. III we derive a new set of inequalities, not
present in the two-beam case, and we comment on them. In Sec.IV,
which-way detection schemes are treated, while in Sec. V, we
discuss the optimization problem for a three-beam example. Sec. VI
is devoted to our concluding remarks.

\section{Visibility and Predictability.}
\setcounter{equation}{0}

We consider an $n$-beam interferometer, in which a beam splitter
splits first a beam of quantum objects ("quantons", in brief) into
$n$ beams, that afterwards converge on a second beam splitter,
where they interfere, giving rise to $n$ output beams. We imagine
that, at some instant of time, the (normalized) wave-functions
$|\psi_i>$ $i=1,\dots,n$ for the individual beams are fully
localized in  the region between the two beam-splitters, and are
spatially well separated from each other, so that
$<\psi_i|\psi_j>=\delta_{ij}$. The state of the quanton, in front
of the second beam-splitter, is then described by a density matrix
$\rho$ of the form: \be \rho=\sum_{ij}
\rho_{ij}\;|\psi_i><\psi_j|\;.\label{ista} \ee The diagonal
elements $\rho_{ii}$ represent the populations $\zeta_i$ of the
beams, and obviously they satisfy the condition: \be \sum_i
\zeta_i={\rm Tr}\,\rho=1\;.\ee

The off-diagonal elements of $\rho$, that we shall denote as
$I_{ij}$, are instead related to the probability $I$ of finding a
quanton in one of the $n$ output beams, according to the following
equation:   \be I= \frac{1}{n}\left(1+ \sum_i \sum_{j \neq i}
\;\;e^{i(\phi_i - \phi_j)}\; I_{ij}\right) \;.\label{prob} \ee
Here, $\phi_i-\phi_j$ is the relative phase between beams $i$ and
$j$. In this paper we consider experimental settings, such that
all these relative phases can be adjustable at will. However, this
is not the case in a number of experimental settings, where the
features of the apparatus may lead to relations among the relative
phases of the beams. When this happens, the output beam intensity
Eq.(\ref{prob}) may be rewritten, by expressing the relative
phases in terms of the independently adjustable ones. An analysis
of complementarity tailored on specific experimental settings,
involving definite relations among the phases, may turn out to be
interesting and useful. However, the purpose of the present paper
is to study the problems arising  when the full freedom allowed by
an $n$-beam setting is taken into account.

Going back to Eq.(\ref{prob}), one notices that $I$ does not
depend at all on the populations $\zeta_i$. In the standard case
of an interferometer with two beams of interfering quantons, a
typical measure of the fringe contrast is the traditional
visibility ${\cal V}$, defined as: \be {\cal V}\equiv \frac{I_{\rm
max}-I_{\rm min}}{I_{\rm max}+ I_{\rm min}}\;, \label{visold}\ee
where $I_{\rm max}$ and $I_{\rm min}$ are, respectively, the
maximum and minimum of $I$. It is easy to verify, using
Eq.(\ref{prob}) with $n=2$, that \be {\cal
V}\,=\,2\,|I_{12}|\;.\label{vistwo} \ee A few years ago,
Greenberger and YaSin \cite{gree} noticed that the general rules
of Quantum Mechanics imply the existence of a simple relation
connecting the visibility ${\cal V}$, to the populations $\zeta_i$
of the beams. They considered the so-called predictability  \be
{\cal P}:=|\zeta_{1}-\zeta_{2}|\;,\label{gree} \ee which can be
interpreted as the a-priori probability for "guessing the way
right", when one has unequal populations of the beams. It is easy
to verify that the general condition \be{\rm Tr} \rho^2 \le
1\;,\label{trace}\ee  turns into the following inequality \be
{\cal V}^2+{\cal P}^2 \;\le\; 1\;.\label{greya} \ee  When it is
saturated, namely for pure states, one can recognize in
Eq.(\ref{greya}) a statement of wave-particle duality, because
then a large predictability of the way followed by the quantons,
implies a small visibility of the interference fringes, and
viceversa.

Independently on any interpretation, the inequality (\ref{greya})
represents a testable relation between measurable quantities, that
follows from the first principles of Quantum Mechanics. Indeed,
the experiments with asymmetric beams of neutrons made by Rauch et
al.\cite{rauc} are compatible with it. It is interesting to
observe that Eq.(\ref{greya}) provides also an operative,
quantitative way to determine how far the beam is from being pure.

One may ask whether an   inequality  analogous to Eq.(\ref{greya})
holds in the multibeam case. Here, one's first attitude would be
to keep the definition of visibility, Eq.(\ref{vistwo}),
unaltered. However, this choice has a severe fault, as we now
explain. Suppose that the beams are made interact with another
system, that we call environment, and assume that the interaction
does not alter the populations of the beams. If the interaction is
described as a scattering process,   its effect is to give rise to
an entanglement of the beams with the environment, such that: \be
|\chi_0><\chi_0|\otimes \rho \rightarrow \rho_{\rm b \&
e}=\sum_{ij} \rho_{ij}\;|\chi_i\!><\chi_j|\otimes
\,|\psi_i\!><\psi_j|\;.\label{deten}\ee Here, $|\chi_0>$ and
$|\chi_i>$ are normalized environments' states (we have assumed
for simplicity that the initial state $|\chi_0>$ of the
environment is pure, but taking a mixture would not change the
result). The entanglement with the environment alters the
probability of finding  a quanton in the chosen output beam.
Indeed, the state $\rho '$ of the beams, after the interaction
with the environment, is obtained by tracing out the environment's
degree of freedom from Eq.(\ref{deten}): \be \rho\,'=\sum_{ij}
\rho_{ij}<\chi_j|\chi_i>\;|\psi_i><\psi_j|\;.\label{rhop} \ee By
plugging $\rho '$ into Eq.(\ref{prob}), we obtain the new
expression for the probability $I\,'$ of finding a quanton in the
selected output beam: \be I\,'= \frac{1}{n}\left(1+ \sum_i \sum_{j
\neq i} \;\;e^{i(\phi_i - \phi_j)}\;
I_{ij}<\chi_i|\,\chi_j>\right)\;.\label{ipri}\ee If we agree that
the visibility $V$ should be fully determined   by the intensity
of the output beam $I'$, we require that it should be defined in
such a way that, for any choice of the environments states
$|\chi_i>$, $V' \le V$. It is easy to convince oneself that the
standard visibility ${\cal V}$ fulfills this requirement for
two-beams, while it does not for a larger number of beams. Indeed,
for two beams, $V' \le V$ is a direct consequence of
Eq.(\ref{vistwo}).  Things are different already with three beams.
Consider for example the   three-beam state, described by
the following density matrix  $\rho$  \be  \rho = \frac{1}{3} \left(%
\begin{array}{ccc}
 1 & -\lambda & \lambda \\
  -\lambda & 1 & -\lambda \\
  \lambda & -\lambda & 1 \\
\end{array}%
\right)\;.\ee It can be checked that $\rho$ is positive definite
if $0 \le \lambda < 1$. A direct computation of the visibility
${\cal V}$, for $\lambda >0$,  gives the result: \be {\cal
V}=\frac{3 \lambda}{2+ \lambda}\;.\ee Suppose now that the
interaction with the environment is such that the environment's
states in Eq.(\ref{deten}) satisfy the conditions:
$|\chi_1>=|\chi_2>$ and $<\chi_1|\chi_3>=<\chi_2|\chi_3>=0$. This
condition is typically realized if the environment interacts only
with the third beam, as it happens, for example, if one scatters
light off the third beam only. This is precisely the type of
situation that is realized, in a four beam context, in the
experiment of Ref.\cite{mei}. With this choice for the states
$|\chi_i>$, the density matrix $\rho\,'$ in
Eq.(\ref{rhop}) becomes: \be \rho\,'=\frac{1}{3} \left(%
\begin{array}{ccc}
 1 & -\lambda & 0 \\
  -\lambda & 1 & 0 \\
 0& 0 & 1 \\
\end{array}%
\right)\;.\ee It can be verified that the new value of the
visibility ${\cal V}\,'$ is: \be {\cal
V}\,'=\frac{4}{3}\lambda\;.\ee We see that, for $1/4 <\lambda<1$,
${\cal V}\,'>{\cal V}$. We believe that these considerations lead
one to abandon ${\cal V}$ as a good measure of the visibility, in
the multibeam case, and to search for a different definition.

Thus we need multibeam generalizations of the above definitions
for the visibility and the predictability. Of course, this is a
matter of choice, but it is clear that the choices for the
definitions of the two quantities are tied to each other, if they
are eventually to satisfy an inequality like Eq.(\ref{greya}).
Indeed a simple reasoning provides us with a possible answer. One
observes that, for any  number of beams, it is still true that
${\rm Tr}\rho^2 \le 1$. Upon expanding the trace, one can rewrite
this condition as: \be \sum_{i}\zeta_{i}^2 \,+\,\sum_i \sum_{j\neq
i}\, |I_{ij}|^2 \le 1\;. \label{obvi} \ee One observes now that
the first sum depends only on the populations $\zeta_{i}$ of the
beams, which should determine the predictability, while the second
sum depends only on the non diagonal elements of $\rho$, which are
the ones that appear in the expression of the intensity $I$ of the
output beam, Eq.(\ref{prob}), and thus determine the features of
the interference pattern. Eq.(\ref{obvi}) suggests that we define
the generalized visibility $V$ as: \be V^2=C\,\sum_i \sum_{j\neq
i}\,|I_{ij}|^2\;, \ee where $C$ is a constant, chosen such that
the range of values of $V$ is the interval $[0,1]$. One finds
$C=n/(n-1)$, and so we get: \be V=\,\sqrt{\frac{n}{n-1}\sum_i
\sum_{j\neq i}\,|I_{ij}|^2}\;,\label{visib} \ee which is the
choice made in \cite{durr}.  It is clear that this definition of
$V$ satisfies the above requirement, that any interaction with the
environment should make $V$ decrease, because, according to
Eq.(\ref{deten}), the moduli $|I_{ij}|^2$ can never get larger, as
a result of the interaction with the environment. Moreover, we see
that for two beams $V=2|I_{12}|$, which coincides with
Eq.(\ref{vistwo}), and so $V={\cal V}$. It is easy to check that
$V$ can be expressed also as  a rms average, over all possible
values of the phases $\phi_i$, of the deviation of the intensity
$I$ of the output beam from its mean value: \be
V=\sqrt{\frac{n^3}{n-1} <(\Delta I)^2>_{\phi}}\;. \label{visibu}
\ee Here the, bracket $< \,
>_{\phi}$ denotes an average with respect to the phases $\phi_i$
and $\Delta I= I-<I>_{\phi}$.

One proceeds in a similar manner with the generalized
predictability $P$. Eq.(\ref{obvi}) suggests that we define $P$
as: \be P^2= A \sum_i \zeta_{i}^2 + B\;, \ee where the constants
$A$ and $B$ should be chosen such that the range of values of
$P^2$ coincides with the interval $[0,1]$. It is easy to convince
oneself that this requirement uniquely fixes $A=n/(n-1)$,
$B=-1/n$, and so we obtain: \be
P=\sqrt{\frac{n}{n-1}\left(-\frac{1}{n}+\sum_i\zeta_{i}^2\right)}\;,\label{genpre}
\ee which is the choice of \cite{durr}. It is easy to check that
this expression coincides with ${\cal P}$, Eq.(\ref{gree}), when
$n=2$. One may
observe that this definition enjoys the following  nice features:\\
\noindent
i) $P$ reaches its maximum value if and only if either one of the
populations $\zeta_i$ is equal to one, and the others are zero, which
corresponds to full predictability of the path;\\
\noindent
ii) $P$ reaches its minimum if and only if all the populations are equal to
each other, which means total absence of predictability;\\
\noindent
ii) $P$ and $P^2$ are strictly convex functions.   This means that, for any
choice of two  sets of populations
$\vec{\zeta}^{\,\prime}=({\zeta}_{1}^{\prime}, \dots,{\zeta}_{n}^{\prime}
)$, and $\vec{\zeta}^{\,\prime\prime}=({\zeta}_{1}^{\prime\prime},
\dots,{\zeta}_{n}^{\prime\prime} )$ and for any $\lambda
\in [0,1]$ one has:
\be
P(\lambda \vec{\zeta}^{\,\prime}+(1-\lambda)\vec{\zeta}^{\,\prime\prime})
\le \lambda P( \vec{\zeta}^{\,\prime}   )+
(1-\lambda)P( \vec{\zeta}^{\,\prime\prime}   )\;,\label{conc}
\ee
where the equality sign   holds if and only if the vectors
$\vec{\zeta}^{\,\prime}$ and $\vec{\zeta}^{\,\prime\prime}$ coincide. A
similar equation holds for $P^2$. This is an important property, because it
means that the predictability (or its square) of any convex combination of
states is never larger than the convex sum of the corresponding
predictabilities (or their squares).\\
\noindent \noindent One can check now that $P^2$ and $V^2$ satisfy
an inequality analogous to Eq.(\ref{greya}): \be V^2\,+\,P^2
\le 1\,,
\label{duine} \ee where the equal sign holds if and only if the
state is pure. This result deserves a number of comments:\\
\noindent 1) As in the two beams case, the above inequality
provides a testable relation between measurable quantities, and it
would be interesting to verify it.\\
\noindent 2) On the level of interpretations, when saturated,
Eq.(\ref{duine}) can be regarded as a statement of wave-particle
duality, in analogy with the two-beam relation, Eq.(\ref{greya}).
In fact, since the quantity $P$ depends only on the populations
$\zeta_i$, $P$ may be interpreted as a particlelike attribute of
the quantons.  On the other side, since the quantity $V$ depends
only on the numbers $I_{ij}$, that determine the interference
terms in the expression of $I$, it is legitimate to regard $V$ as
a measure of the wavelike attributes
of the quanton.\\
\noindent 3) However, the quantity $P$ does not carry the same
meaning as the quantity ${\cal P}$ used in the two-beam case, and
the name "predictability" given to it in Ref.\cite{durr}   is not
the most appropriate. Indeed, from the point of view of
statistical decision theory \cite{helstrom}, the natural
definition of predictability would not be that in
Eq.(\ref{genpre}), bur rather the following.  If one interprets
the number $\zeta_i$  as the probability for a quanton to be in
the beam $i$, and if  one decides to bet every time on the most
populated beam $\bar i$, the sum $\sum_{i \neq \bar i}\zeta_i$
represents the probability of loosing the bet. Then, it is natural
to define the predictability ${\cal P}_n$ as: \be {\cal P}_n=1-
\frac{n}{n-1}\sum_{i \neq \bar i} \zeta_i \;\label{gregen},\ee
where the normalization is fixed by the requirement that  ${\cal
P}_n=0$, if the beams are equally populated, and ${\cal P}_n=1$,
if any of the populations is equal to one. For $n=2$, this
definition reduces to that used by Greenberger and YaSin, in
Eq.(\ref{gree}), and in fact it was proposed as a generalization
of it in Ref.\cite{jaeg}. It is surely possible to write
inequalities involving ${\cal P}_n$ and $V$, but, as far as we
know, none of them is saturated by arbitrary pure states,
differently from Eq.(\ref{duine}). So, one is faced with a
situation in which the less intuitive notion of "predictability",
given by Eq.(\ref{genpre}), enters in a sharp relation with the
visibility, while the most intuitive one, given by
Eq.(\ref{gregen}), enters in a relation with the visibility, that
is not saturated even for pure states.

\section{Higher order inequalities.}
\setcounter{equation}{0}

In a multibeam interferometer a new interesting  feature is
present, which is absent in the two-beam case, and puts
Eq.(\ref{greya}) into a new perspective. In fact,
Eq.(\ref{duine}), that relates the populations of the beams
$\zeta_i$ to the features of the interference  fringes, is only
the first of a collection of inequalities, that we now discuss.
The new inequalities, exactly like Eq.(\ref{duine}), rest on the
first principles of Quantum Mechanics and can be derived along
similar lines, by considering higher powers of the density matrix
$\rho$. Indeed, for $n$ beams, one has the following $n-1$
independent inequalities: \be {\rm Tr} \;\rho^m \,\le
\,1\;\;\;\;\;\;m=2, \dots,n\;.\label{ours} \ee For example, with
three beams, if we take $m=3$ we obtain: \be 0 < \sum_i \zeta_i^3
+ 3 \sum_i \zeta_i \sum_{j \neq i} |I_{ij}|^2 +
3\,(I_{12}I_{23}I_{31}+ {\rm h.c.}) \le 1 \;.\label{inthr} \ee
This inequality, like Eq.(\ref{duine}), may be translated in terms
of physically measurable quantities, although in a more elaborate
way. First, we notice that the combination of non-diagonal
elements of the density matrix, that appears in the last term of
the r.h.s. of the above Equation represents the third moment of
the intensity $I$ of the output beam: \be (I_{12}I_{23}I_{31}+
{\rm h.c.})=\frac{<(\Delta I)^3>_{\phi}}{<I>_{\phi}^3}\;. \ee On
the other side, the quantities $|I_{ij}|^2$ that appear in the
middle terms, are related, as in Eq.(\ref{vistwo}), to the
visibilities $V_{ij}$ of the three interference patterns, that are
obtained by letting the beams $i$ and $j$ interfere with each
other, after intercepting the remaining beam. Therefore, we  may
rewrite Eq.(\ref{inthr}) as: \be  0 < \sum_i \zeta_i^3 +
\frac{3}{4} \sum_i \zeta_i \sum_{j \neq i} V_{ij}^2 + 3\,
\frac{<(\Delta I)^3>_{\phi}}{<I>_{\phi}^3}\le 1 \;,\ee which shows
clearly that the novel inequality is a testable relation, to be
checked by experiment.

This example illustrates the general structure of the new higher
order inequalities. As the number $n$ of beams and the power of
$m$ in Eq.(\ref{ours}) increase, higher and higher moments of the
intensity $I$ will appear. Furthermore, data related to the
interference patterns formed by  all possible subsets of beams
that can be sorted out of the $n$ beams, will appear.

A few comments are in order. On one side, the higher order
inequalities are similar to Eq.(\ref{duine}), in that they are all
testable in principle, and become equalities for beams in a pure
state. On the other side, differently from Eq.(\ref{duine}), they
do not exhibit a natural splitting of the particlelike quantities
$\zeta_i$ from the wavelike quantities $I_{ij}$, into two
separate, positive definite terms.

The existence of this sequence of inequalities suggests that, from
the point of view of complementarity, the two-beam and the
multibeam case are different. For two-beams, the basic properties
of the density matrix are completely expressed in terms of a
single duality relation, like Eq.(\ref{greya}). In the multibeam
case, a whole sequence of independent inequalities is needed, if
one is to fully express the basic properties of the density
matrix. Except for the first one, none of  these inequalities
seems to be related in any simple way to the intuitive concept of
wave-particle duality. It seems than that the lowest-order
inequality, Eq.(\ref{duine}), still carries an idea of
wave-particle duality, but only at the cost of averaging out the
effects related to higher order moments.

\section{Which-way detection.}
\setcounter{equation}{0} The notion of predictability, introduced
in Sec.II, does not express any real knowledge of the path
followed by individual quantons, but at most our a-priori ability
of predicting it. A more interesting situation arises if  the
experimenter actually tries to gain which-way information on
individual quantons, by letting them interact with a detector,
placed in the region where the beams are still spatially
separated. The analysis proceeds assuming that the detector also
can be treated as a quantum system, and that the particle-detector
interaction is described by some unitary process. A detector can
be considered as a part of the environment, whose state and whose
interaction with the beams can, to some extent, be controlled by
the experimenter. If we let $|\chi_0>$ be the initial state of the
detector (which we assume to be pure, for simplicity), the
interaction with the particle will give rise to an entangled
density matrix ${\rho}_{\rm b \& e}$, of the form considered
earlier, in Eq.(\ref{deten}). This time, however, we interpret the
states  $|\chi_i\!>$ as $n$ normalized (but not necessarily
orthogonal !) states of the which-way detectors. The existence of
a correlation between the detector state $|\chi_i>$ and the beam
$|\psi_i>$, in Eq.(\ref{deten}), is at the basis of the detector's
ability to store which-way information.  We observed earlier that
the very interaction of the quantons with the detector, causes, as
a rule, a decrease in the visibility. According to the intuitive
idea of the wave-particle duality, one would like to  explain this
decrease of the visibility  as a consequence of the fact that  one
is trying to gain which-way information on the quantons. In order
to see if this is the case, we need read out the which-way
information stored in the detector. We thus consider the final
detector state $\rho_D$, obtained by taking a trace of
Eq.(\ref{deten}) over the particle's degrees of freedom: \be
\rho_D= \sum_{i} \zeta_i \;|\chi_i\!><\chi_i| \;. \ee As we see,
$\rho_D$ is  a mixture of the $n$ final states $|\chi_i\!>$,
corresponding to the $n$ possible paths, weighted by the fraction
$\zeta_i$ of quantons taking the respective path. Thus the problem
of determining the trajectory of the particle reduces to the
following one: after the passage of each particle, is there a way
to decide in which of the $n$ states $|\chi_i\!>$ the detector was
left?  If the states $|\chi_i\!>$ are orthogonal to each other,
the answer is obviously yes.  If, however, the states $|\chi_i\!>$
are not orthogonal to each other, there is no way to unambiguously
infer the path: whichever detector observable $W$ one picks, there
will be at least one eigenvector of $W$, having a non-zero
projection onto more than one state $|\chi_i\!>$. Therefore, when
the corresponding eigenvalue is obtained as the result of a
measurement, no unique detector-state can be inferred, and only
probabilistic judgments can be made. Under such circumstances, the
best  the experimenter can do is to select the observable that
provides as much information as possible, ${\it on \; the
\;average}$, namely after many repetitions of the experiment. Of
course, this presupposes the choice of a definite criterion to
measure the average amount of which-way information delivered by a
certain observable $W$.\\
\noindent Let us see in detail how this is done. Consider an
observable $W$, and let $\Pi_{\mu}$ the projector onto the
subspace of the detector's Hilbert space ${\cal H}_D$, associated
with the eigenvalue $w_{\mu}$. The a-priori probability $p_{\mu}$
of getting the result $w_{\mu}$ is: \be p_{\mu}={\rm
Tr}_D\,(\Pi_{\mu}\,\rho_D)\;=\sum_{i}\zeta_i \, P_{i
\mu}\;,\label{pmuo} \ee where ${\rm Tr}_D$ denotes a trace over
the detector's Hilbert space ${\cal H}_D$ and $P_{i
\mu}=|<\chi_i|\Pi_{\mu}|\chi_i>|^2$.  The quantity $\zeta_i \,P_{i
\mu}$ coincides with the probability of getting the value
$w_{\mu}$, when all the beams, except the $i$-th one, are
intercepted before reaching the detector, and indeed this provides
us a way to measure the numbers $\zeta_i\,P_{i \mu}$. When the
interferometer is operated with $n$-beams, one may interpret the
normalized probabilities $Q_{i\mu}$: \be
Q_{i\mu}=\frac{\zeta_i\,P_{i\mu}}{p_{\mu}} \ee as the a-posteriori
relative probability, for a particle to be in the $i$-th beam,
provided that the measurement of $W$ gave the outcome
$w_{\mu}$.\\
\noindent On the other side, if $W$ is measured after the passage
of each  quanton, one can sort the quantons in the output beam
into distinct subensembles, according to the result $w_{\mu}$ of
the measurement.   The subensembles of quantons are described by
density matrices $\rho_{(\mu)}$ of the form: \be
\rho_{(\mu)}=\frac{1}{p_{\mu}} \,{\rm Tr}_D\, (\Pi_{\mu}\,
{\rho}_{\rm b \& e}):=\sum_{ij}
\rho_{(\mu)ij}|\psi_i><\psi_j|\;,\label{suben} \ee where we
defined:
\be\rho_{(\mu)ij}=\frac{1}{p_{\mu}}<\chi_j|\Pi_{\mu}|\chi_i>\,\rho_{ij}\;.\ee
We see that the a posteriori probabilities $Q_{i \mu}$ coincide
with the diagonal elements of the density matrices
$\rho_{(\mu)ij}$, and thus represent also
the populations of the beams, for the sorted subensembles of quantons.\\
\noindent Let us consider now the case of two beams. For each
outcome $w_{\mu}$, one can consider the  predictability ${\cal
P}_{\mu}(W)$ and the visibility ${\cal V}_{\mu}(W)$, associated
with the corresponding subensemble of quantons: \be {\cal
P}_{\mu}(W)=|\rho_{(\mu)11}-\rho_{(\mu)22}| =| Q_{1\mu}- \,Q_{2
\mu}|\;, \ee \be {\cal V}_{\mu}(W)= 2 \,|\rho_{(\mu)12}|\;.\ee
Notice that both quantities depend, of course, on the observable
$W$.  It is clear that an inequality like Eq.(\ref{gree}) holds
for each subensemble, separately: \be {\cal P}^2_{\mu}(W)+{\cal
V}^2_{\mu}(W) \le 1\;.\label{insub}\ee The equality sign holds if
and only if the subensemble is a pure state, which is surely the
case if the beams and the detector are separately prepared in pure
states, before they interact. When the eigenvalue $w_{\mu}$ is
observed, it is natural to define the average amount  of which-way
knowledge delivered by $W$ as the predictability ${\cal
P}_{\mu}(W)$ of the corresponding subensemble of quantons.  In
order to measure the overall ability of the observable $W$ to
discriminate the paths, one defines a quantity ${\cal K}(W)$
\footnote{Indeed, Englert considers the "{\it likelihood} ${\cal
L}_W$ for guessing the way right. In our notation, ${\cal
L}_W=(1+{\cal K}(W))/2$.}, which is some average of the partial
predictabilities ${\cal P}_{\mu}(W)$. The procedure implicitly
adopted by Englert in \cite{englert}, is to define ${\cal K}(W)$
as the weighted average of the  numbers ${\cal P}_{\mu}(W)$, with
weights provided by the a priori probabilities $p_{\mu}$: \be
{\cal K}(W)= \sum_{\mu} p_{\mu} \,{\cal
P}_{\mu}(W)\;.\label{engtwo} \ee One can introduce also the
"erasure visibility" \cite{scully2}, relative to $W$, as the
weighted average of the partial visibilities: \be {\cal
V}(W)=\sum_{\mu}p_{\mu}\,{\cal V}_{\mu}(W)\;.\ee For any $W$,
these quantities can be shown to satisfy the following inequality,
that is a direct consequence of Eq.(\ref{insub}): \be {\cal
K}^2(W)+{\cal V}^2(W)  \le 1\;.\label{inew}\ee  Moreover, one can
prove that: \be {\cal P}^2 \le {\cal K}^2(W)\;,\ee which gives
expression to the intuitive idea that any observable $W$, that we
decide to measure, provides us with a better knowledge of the
path, than that available on the basis of a mere a priori
judgement. One has also the other inequality \be {\cal V}^2 \le
{\cal V}^2(W)\;.\label{vvw}\ee For the proofs of these
inequalities, we address the reader to Ref.(\cite{berg}), where
they are derived in a number of independent ways.
 In the
so-called which-way sorting schemes, it is natural to select the
observable $W$ such as to maximize ${\cal K}(W)$, and one then
defines the distinguishability ${\cal D}$ of the paths   as the
maximum value of ${\cal K}(W)$: \be {\cal D}= \max_W \{{\cal
K}(W)\}\;.\label{engthr} \ee  It is easy to see that
Eqs.(\ref{inew}), (\ref{vvw}) and (\ref{engthr}) together imply
the following inequality, analogous to Eq.(\ref{greya}), first
derived by Englert in Ref.\cite{englert}: \be {\cal D}^2\,+\,{\cal
V}^2 \le 1\;.\label{englert} \ee Thus, given the visibility ${\cal
V}$, there is an upper bound for the distinguishability, set by
the above relation. But Englert in fact proves much more than
this: he shows that Eq.(\ref{englert}) becomes an {\it identity},
when both the beams and the detector are in a pure state. In our
opinion, this fact is essential to justify the interpretation of
Eq.(\ref{englert}) as a statement of the complementary character
of the wave and  particle attributes of a  quanton. In fact, this
implies that, when the beam of quantons and the detector are as
noiseless as they can possibly be in Quantum Mechanics, namely
when they are in pure states, an increase in any of the two terms
is necessarily accompanied by an exactly quantifiable
corresponding
decrease of the other.  \\
\noindent A possible generalization of the above considerations,
to the multibeam case, is as follows \cite{durr}. One sorts again
the quantons, into subensembles, depending on the outcome of the
measurement of $W$. For each outcome $w_{\mu}$, one uses the
generalized predictability $P$ in Eq.(\ref{genpre}), and the
generalized visibility $V$ in Eq.(\ref{visib}), to define the
"conditioned which-way knowledge"
  $K_{\mu}(W)$:
\be K_{\mu}(W)=\sqrt{\frac{n}{n-1}\left(-\frac{1}{n}+\sum_i\  Q_{i
\mu}^2\right)}\;, \label{conww} \ee and the "partial erasure
visibility" $V_{\mu}(W)$: \be
V_{\mu}(W)=\,\sqrt{\frac{n}{n-1}\sum_i \sum_{j\neq
i}\,|\rho_{(\mu)ij}|^2}\;.\label{pvisib}\ee In view of
Eq.(\ref{duine}), they satisfy an inequality analogous to
Eq.(\ref{insub}): \be K_{\mu}^2(W)+V_{\mu}^2(W)\le 1\;.
\label{mfun}\ee Again, as in the two beam case, the equality sign
holds if the subensembles are pure. The author of Ref.\cite{durr}
considers now two different definitions for the "which-way
knowledge" and the "erasure visibility", associated to $W$, as a
whole. The first one is closer to Eq.(\ref{engtwo}): \be K(W):=
\sum_{\mu} p_{\mu}\, K_{\mu}(W)\;\;,\;\;\;\;\; V(W):= \sum_{\mu}
p_{\mu}\, V_{\mu}(W).\label{kw} \ee The second one, inspired by
the work of Brukner and Zeilinger \cite{bruk}, is\footnote{We use
here a notation different from that of Ref.\cite{durr}. Our
$\tilde{K}^2(W)$ and  $\tilde V^2(W)$ correspond, respectively, to
$n/(n-1) I_{KW}$ and $n/(n-1) I_{VW}$, in \cite{durr}. }: \be
\tilde{K}^2(W):=\sum_{\mu} p_{\mu}\,
K_{\mu}^2(W)\;\;,\;\;\;\;\;\tilde{V}^2(W):=\sum_{\mu} p_{\mu}\,
V_{\mu}^2(W).\label{bz} \ee The quantities introduced above, are
related by   the following chains of inequalities, the proofs of
which can be found in \cite{durr}: \be V \le V(W) \le \tilde
{V}(W)\;,\;\;\;P \le K(W) \le \tilde{K}(W)\;.\label{chai}\ee These
inequalities show that $\tilde{K}(W)$ and $\tilde{V}(W)$ provide
more efficient measures for the average which-way information, and
for the erasure visibility, respectively. However, the author of
Ref.\cite{durr} observes that the quantities $K(W)$  and $V(W)$
are preferable to ${\tilde K}(W)$, and $\tilde{V}(W)$,
respectively,  because they are the ones that reduce, for $n=2$,
to the definitions used in the two-beam case. We would like to
point out that, since $K_{\mu}(W)$ and $V_{\mu}(W)$ are
essentially variances of the diagonal and non-diagonal elements,
respectively, of the density matrices for the subensembles of
quantons, it appears more natural, from a statistical point of
view, to combine them in quadrature, as done in Eq.(\ref{bz}).
This suggests that one should  adopt the definition with the
quadrature also in the two-beam case.

By taking the suprema   of all the  quantities defined above, over
all possible observables $W$, one can define a set of  four
 quantities, that
characterize the state $\rho$ of the beams. For example, upon
taking the  maxima of $K(W)$ and $\tilde{K}(W)$, we end up with
two possible definitions for the which-way distinguishability, $D$
and $\tilde D$, respectively: \be D=\max_W\{K(W)\}\;\;,\;\;\;\;
{\tilde D} = \max_W \{\tilde{K}(W)\}\;.\label{tildd} \ee
Similarly, by taking the suprema of $V(W)$ and $\tilde{V}(W)$, we
obtain two definitions of the so-called "coherence" of the beams
\cite{meystre}: \be C=\sup_W\{V(W)\}\;\;,\;\;\;\;  {\tilde C} =
\sup_W \{\tilde{V}(W)\}\;.  \ee (The reader may found in
Ref.\cite{berg} an explanation of why one has maxima, in the
definition of distinguishability, and only suprema in that of
coherence.)

The quantities introduced above, satisfy a set of inequalities,
that all follow from the chains of inequalities Eqs.(\ref{chai}),
and from the following inequality, that can be obtained  from
Eq.(\ref{mfun}), on averaging over all possible outcomes
$w_{\mu}$: \be \tilde{K}^2(W)+\tilde{V}^2(W)\le 1\;. \label{mw}\ee
It is clear that this inequality is saturated, regardless of the
observable $W$,  when the state of the combined  detector-beam
system is pure. This is an immediate consequence of
Eq.(\ref{mfun}).

One of the central results of Ref.\cite{durr} is the following
inequality, generalizing Eq.(\ref{englert}): \be {\tilde
D}^2\,+\,V^2 \le 1\;.\label{tildein} \ee Since ${\tilde D} \ge D$,
this also implies: \be {  D}^2\,+\,V^2 \le \, 1\;.\label{durrin}
\ee Thus we see that also in the multibeam case, the visibility
$V$ sets an upper limit for the amount of which-way information,
irrespective of how one measures it, via $D$ or $\tilde D$.  In
Ref.\cite{durr} it is suggested that the above two inequalities
provide multibeam generalizations of the two-beam wave-particle
duality relation Eq.(\ref{englert}).

Even if Eq.(\ref{durrin}) and Eq.(\ref{tildein}) represent correct
inequalities, that can be tested in an experiment, in our opinion,
their interpretation as an expression of wave-particle duality
appears disputable. The root of the problem is that the above
inequalities, differently from the two beam case, cannot be
saturated, in general, even  if the beams and the detector are
prepared in pure states (in Appendix I, we actually prove that
Eq.(\ref{durrin}), for example, can be saturated  only if $D=P$,
which means that the detector does not provide any information).
Therefore, one may conceive the possibility of designing two
which-way detectors $D_1$ and $D_2$, such that $V_1
> V_2$, while, at the same time, $D_1 > D_2$. This  possibility,
which conflicts with the intuitive idea of complementarity,
actually occurs, as we  anticipated in Ref.\cite{bimonte}, and as
we report in the next Section.

\section{A three-beam  example.}
\setcounter{equation}{0}

The example discussed in Ref.\cite{bimonte},  was based on a three
beam interferometer with equally populated beams, described by the
pure state: \be \rho=\frac{1}{3}\sum_{ij} \;|\psi_i><\psi_j|\;.
\ee For the sake of simplicity, it was assumed there that the
detector's Hilbert space was a two-dimensional space ${\cal H}_2$.
Its rays were described via the Bloch parametrization, such that:
\be \frac{1+\hat n\, {\cdot} \,\vec{\sigma}}{2}
=|\chi><\chi|\;,\label{block} \ee where $\hat n$ is a unit
three-vector and $\vec{\sigma}=(\sigma_x, \sigma_y, \sigma_z)$ is
any representation of the Pauli matrices.  We denoted by
$|\hat{n}><\hat{n}|$ the ray corresponding to the vector $\hat n$.
We required that the directions $\hat{n}_+,\hat{n}_-,\hat{n}_0$,
associated with the states $|\chi_i>$, were coplanar, and such
that $\hat{n}_+$ and $\hat{n}_-$ both formed an angle $\theta$
with $\hat{n}_0$ We imagined that $\theta$ could be varied at
will, by acting on the detector, and in Ref.\cite{bimonte} we
obtained the following expressions for the visibility $V$ and the
distinguishability $D$, as functions of $\theta$: \be V(\theta)=
\sqrt{\frac{1+\cos\theta+\cos^2 \theta}{3}}\;,\label{visex} \ee
\be \;\;\,D(\theta)=\,\frac{1}{\sqrt 3} \,\sin \theta
\;\;\;\;\;\;\;\;\;\;\;{\rm for}\;\;\;\;\;  0 \le \theta \le 2/3\,
\pi\;,\label{smal} \ee \be D(\theta)=\frac{2}{3} \sin^2
\left(\frac{\theta}{2}\right) \;\;\;\;\;\;\;\;{\rm for}\;\;\;
2/3\,\pi \le \theta \le \pi\;.\label{lar} \ee The values of $V$
and $D$ are plotted in the figure. By looking at it, one realizes
that something unexpected happens:  while in the interval $0 \le
\theta < \pi/2$, $V$ decreases and $D$ increases, as expected from
the wave-particle duality, we see that in the interval $\pi/2 \le
\theta \le \pi$, $V$ and $D$ decrease and increase simultaneously!
We see that if we pick two values $\theta_1$ and $\theta_2$ in
this region, we obtain two which-way detectors, that precisely
realize the situation described at the end of the previous
Section.

The analysis of Ref.\cite{bimonte}, that we have summarized here,
is not realistic though, because of the simplifying assumption of
a detector with a two-dimensional Hilbert space of states. Even
assuming that the detector's final states $|\chi_i>$ span a
two-dimensional subspace ${\cal H}_2$, still one has to take into
account that the full Hilbert space ${\cal H}_D$ of a realistic
device is infinite-dimensional. Now, it is known from the theory
of quantum detection \cite{helstrom, peres} that the optimum
discrimination among an assigned set of quantum states, is not
always achieved by an observable that leaves invariant the
subspace spanned by them. However, the value of $D$ quoted above
corresponds to maximizing the which-way knowledge over the
restricted set of detector's observables $W$, that leave invariant
the subspace ${\cal H}_2$. Then, in order to complete the proof,
we need to show that no observable in ${\cal H}_D$ can perform
better than the one determined in Ref.\cite{bimonte}, by
considering only operators that live in ${\cal H}_2$.
\begin{figure}[t!]
\epsfig{figure=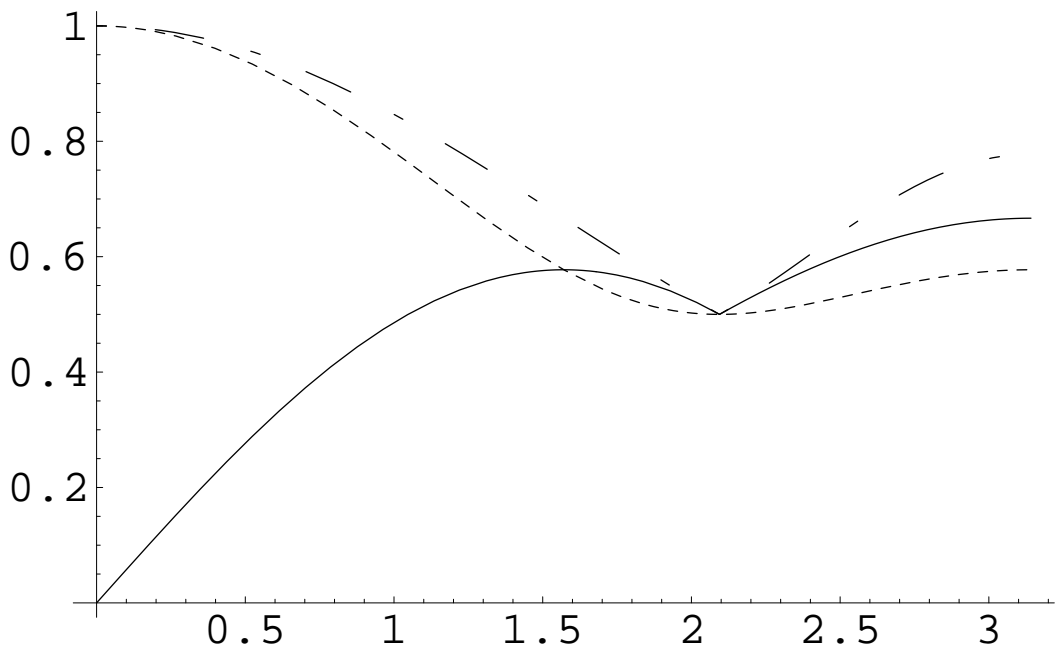,angle=0,width=400pt} \caption{Plots of
the quantities $D$ (solid line), $V$ (dotted line), and  $D^2+V^2$
(dashed line), as functions of $\theta$ in a three beam
situation.} \label{fig}
\end{figure}
Filling this gap, is by no means an easy job,  because it is a
matter of solving an optimization problem in an
infinite-dimensional Hilbert space. There is no general strategy
for solving this sort of   problems, and we can rely only on few
known general results \cite{helstrom,peres,davies}. The interested
reader can find the lengthy procedure to compute $D$ in Appendix
II. Here, we content ourselves with sketching the method followed,
and presenting the results.

For the sake of definiteness,  let us agree to use $K(W)$ as our
measure of the which way-information. At the end of this Section,
we shall discuss what changes if one instead uses ${\tilde K}(W)$.
The determination of the optimal observable $W_{\rm opt}$ is
facilitated by the observation that, even when ${\cal H}_D$ is
infinite-dimensional, the problem can be formulated entirely in
the subspace ${\cal H}_2$, as we now explain. One observes that
the probabilities $P_{i\mu}$ that enter in the definition of
$K(W)$ can be written also as: \be P_{i\mu}=
<\chi_i|\Pi_{\mu}|\chi_i>=
<\chi_i|\Pi\,\Pi_{\mu}\,\Pi|\chi_i>\equiv
<\chi_i|A_{\mu}|\chi_i>\;,\label{amu} \ee where $\Pi$ is the
orthogonal projector onto ${\cal H}_2$, and
$A_{\mu}=\Pi\,\Pi_{\mu}\,\Pi$ is a positive (hermitian) operator
on the subspace ${\cal H}_2$. Thus we see that the operators
$A_{\mu}$ contain all the information we need, about $W$, in order
to compute the which-way knowledge. It is to be noticed that
$A_{\mu}$ are not projection operators, in general. However, they
must provide a decomposition of the identity onto ${\cal H}_2$,
since: \be \sum_{\mu} A_{\mu}= \sum_{\mu} \Pi\, \Pi_{\mu} \Pi =\Pi
(\sum_{\mu} \Pi_{\mu} )\Pi=\Pi.\label{POVM}\ee Such a collection
of operators on ${\cal H}_2$, provides an example of what is known
in Mathematics as a Positive Operator Valued Measure (POVM in
short). Notice though that, while any hermitian operator in ${\cal
H}_D$ gives rise, by projection, to a POVM in ${\cal H}_2$, the
converse may not be true. \footnote{In effect, this problem arises
only if ${\cal H}_D$ is finite dimensional. If ${\cal H}_D$ is
infinite dimensional, all POVM's are acceptable, because a general
theorem due to Neumark \cite{neum} ensures that all POVM's of any
Hilbert space, can be realized as projections of self-adjoint
operators from a larger Hilbert space.} Our strategy to determine
$W_{\rm opt}$  is then to search  first for the optimal POVM
$A_{\rm opt}$ in ${\cal H}_2$ (the notion of which way knowledge
is obviously defined for an arbitrary POVM, as well), and to check
at the end if   $A_{\rm opt}$ can  be realized by projecting onto
${\cal H}_2$ an operator $W$  in ${\cal H}_D$, as in
Eq.(\ref{amu}). If this is the case, $W$ is guaranteed to be
optimal, and we can say that $D=K(A_{\rm opt})$. The determination
of $A_{\rm opt}$ is facilitated by a general theorem
\cite{davies}, that states that for any measure of the which-way
knowledge that is a weighted average of a convex function, the
optimal POVM consists of rank-one operators. This is the case for
the which-way knowledge $K$, which is a weighted average of the
predictability $P$, which indeed is a convex function. The
$A_{\mu}$ being rank-one operators, we are ensured that there
exist non-negative numbers $2 \,\alpha_{\mu} \le 1$ and unit
vectors $\hat{m}_{\mu}$ such that: \be A_{\mu}=2 \,\alpha_{\mu}
|\hat{m}_{\mu}><\hat{m}_{\mu}|=
\alpha_{\mu}(1+\hat{m}_{\mu}{\cdot}\vec{\sigma})\;.\label{ranuno}\ee
The condition for a POVM, Eq.(\ref{POVM}) is equivalent to the
following conditions, for the numbers $\alpha_{\mu}$ and the
vectors $\hat{m}_{\mu}$: \be \sum_{\mu}
\alpha_{\mu}=1\;\;,\;\;\;\;\;\;\sum_{\mu}\alpha_{\mu}
\hat{m}_{\mu}=0\;. \label{cond} \ee The interested reader may find
in Appendix II how the optimal POVM can be determined. Here we
just report the result: for all values of $\theta$, ${A_{\rm
opt}}$ turns out to have only two non vanishing elements,
$A_{\pm}$,  such that: \be A_{\pm}=
\frac{1\pm\sigma_x}{2}\;\;\;\;\;\;\;{\rm for}\;\;\; 0 \le \theta <
2 \pi/3\;,\label{pvmsm} \ee \be A_{\pm}=
\frac{1\pm\sigma_z}{2}\;\;\;\;\;\;\;{\rm for} \;\;\;2 \pi/3 <
\theta \le \pi \;.\label{opt}\ee  It is clear that the operators
$A_{\pm}$ coincide with the projectors found in
Ref.\cite{bimonte}, showing that it was indeed sufficient to carry
out the optimization procedure in ${\cal H}_2$.

It should be appreciated that this coincidence is by no means
trivial, and strictly depends on the choice of $K(W)$ as a measure
of which-way knowledge. For example, for $\theta=2\pi/3$, it is
known \cite{helstrom, peres}, that, with either Shannon's entropy
or Bayes' cost function as measures of information, the optimal
POVM actually consists of three elements, and thus it is not
associated with an operator in ${\cal H}_2$.

Then, our observation that the inequality Eq.(\ref{duine}) fails
to carry the physical picture associated with the idea of
complementarity is now fully demonstrated. We have checked that a
similar conclusion can be drawn if, rather than $K$, one uses the
alternative definition of distinguishability $\tilde D$ provided
by Eq.(\ref{tildd}). In fact, it turns out that the optimal POVM
for $\tilde K$ coincides with the one found earlier, in the
interval $0 \le \theta < \,2/3\,\pi$, and so $\tilde D=D$. The
proof of this can be found in Appendix II.

\section{Conclusions}

The intuitive concept of Complementarity has found, in the case of
two-beams interference experiments, a satisfactory, fully quantum
mechanical formulation as interferometric duality. In this paper,
we critically analyzed the difficulties encountered in the attempt
of generalizing this concept to multibeam experiments, and
discussed the shortcomings  that are present, in our opinion, in
recent proposals. It seems to us fair to say that interferometric
duality has not yet found a proper formulation, in the multibeam
case. To justify this conclusion, let us recall the different
points we have elaborated in the paper.

In the two-beam case, general quantum mechanical requirements on
the density matrix imply the Greenberger-YaSin  inequality, that,
when saturated, expresses interferometric duality. This inequality
has been generalized  to the multibeam case \cite{durr},  leading
to a formal definition of interferometric duality for more than
two beams. The price payed is that the corresponding generalized
concept of predictability has lost the intuitive connection with
minimizing the error in guessing the way right. The traditional
concept of predictability may enter, together with the generalized
visibility, in an inequality that is not saturated, and then
cannot convey the idea of complementarity, which requires that a
better visibility is necessarily related to a loss in information.

We have shown that general requirements of quantum mechanics imply
new inequalities, that are not present in the two beam case. These
inequalities are again experimentally testable. They deserve
further study but, at the present, they do not seem to exhibit a
direct relation with the idea of complementarity.

Interferometric duality may be fully analyzed  only in the
presence of which-way detectors. In the two beam case, Englert has
shown that the visibility enters, with the distinguishability,
into an inequality, that is saturated for pure states. As
maximizing the distinguishability, minimizes the error in guessing
the way right by performing a measurement, this relation fully
expresses interferometric duality. In deriving an analogous
inequality for the multibeam case, D\"{u}rr has introduced two
alternative notions of distinguishability. However, we have shown
that this inequality is never saturated, apart from trivial cases.
Then, a pure inequality may be consistent with a situation in
which an increase (decrease) in visibility goes together with an
increase (decrease) in distinguishability, contrary to the
intuitive idea of interferometric duality. We have given a  full
proof that this possibility actually occurs in a realistic
example. The inequalities proposed by D\"{u}rr, in terms of
generalized visibility and distiguishability,  are then correct
quantum mechanical relations, testable in principle, but they fail
to convey the idea of interferometric duality.

It is seems then fair to conclude that interference duality in
multibeam experiments has not yet been properly formulated. We
leave the problem open, but we notice it is by no means necessary
that quantum mechanics should provide us with an exact formulation
of this concept in the multibeam case. May be, one should content
him(her)self with its formulation in the two beam case, where the
semiclassical intuitive idea of complementarity was first
introduced. May be, Quantum Mechanics provides us just with the
values of observable quantities, and experimentally testable
inequalities. The analysis we have performed may hint in this
direction, but further investigation is required.

\section{Acknowledgments}

We gratefully acknowledge an interesting discussion with Prof. G. Marmo and
Prof. E.C.G. Sudarshan. The work of G.B. was partially supported by the
PRIN {\it S.Inte.Si.}. The work of R.M. was partially supported by EC
program HPRN-EC-2000-00131, and by the PRIN "{\it Teoria dei campi,
superstringhe e gravit\`{a}}".

\section{Appendix I}
\setcounter{equation}{0}

In this Appendix, we prove the following result: for any number
$n>2$ of beams in a pure state $\rho$, and any detector in a pure
initial state, the inequality Eq.(\ref{durrin}) is satisfied as an
equality if and only if $D=P$, namely when the detector provides
no information at all. The proof consists in showing that the
equal sign in Eq.(\ref{durrin}) holds only if the detector states
$|\chi_i>$ are proportional to each other, which obviously implies
$D=P$. \noindent Consider the optimal operator $W_{opt}$ such that
$K(W_{opt})=D$ (we assume that such an operator exists), and let
$V(W_{opt})$ be the corresponding erasure visibility. It follows
then from Eqs.(\ref{chai}) and Eqs.(\ref{mw}) that: \be
D^2\,+\,V^2 \le D^2 + \,V^2(W_{opt}) =
K^2(W_{opt})\,+\,V^2(W_{opt}) \le
 \tilde{K}^2(W_{opt})\,+\,\tilde{V}^2(W_{opt})= 1 \;.
\ee  We see that a necessary condition to have $D^2+V^2=1$ is
that: \be V=V(W_{opt})\;.\label{contwo} \ee In what follows, we
shall not consider the trivial case $V=0$, and we shall suppose
that $V>0$. In order to study Eq.(\ref{contwo}), we take advantage
of the fact that, $K(W)$ being convex, the spectrum of $W_{opt}$
can be taken to be non degenerate \cite{davies}.   If we let
$|w_{\mu}>$ the   eigenvectors of $W_{opt}$, with non-vanishing
projection onto some of the states $|\chi_i>$, by using the
expressions Eq.(\ref{pvisib}) for the partial visibilities, we can
write:
$$
\frac{n-1}{n}\,V^2(W_{opt})=\frac{n-1}{n}\sum_{\mu}\sum_{\nu}
\,p_{\mu} p_{\nu}\, V_{\mu} V_{\nu}=
$$
\be = \sum_{\mu} \sum_{\nu}  \sqrt{\sum_i \sum_{j \neq i}
|{<w_{\mu} |\tilde{\rho}_{ij} |w_{\mu}>} |^2} \sqrt{\sum_p \sum_{q
\neq p}{|<w_{\nu} |\tilde{\rho}_{pq}
|w_{\nu}>}|^2}\;,\label{stepo} \ee where \be \tilde{\rho}_{ij}:=
{\rho}_{ij}\,|\chi_i><\chi_j|\;. \ee Now, the Cauchy-Schwarz
inequality for real vectors implies that: \be \sqrt{\sum_i \sum_{j
\neq i} |{<w_{\mu} |\tilde{\rho}_{ij} |w_{\mu}>} |^2} \sqrt{\sum_p
\sum_{q \neq p}{|<w_{\nu} |\tilde{\rho}_{pq} |w_{\nu}>}|^2}\ge
\sum_i \sum_{j \neq i}|{<w_{\mu} |\tilde{\rho}_{ij}
|w_{\mu}>}|{\cdot} |<w_{\nu} |\tilde{\rho}_{ij}
|w_{\nu}>|\;.\label{cs} \ee Upon using this relation into
Eq.(\ref{stepo}), we obtain: \be \frac{n-1}{n}\;V^2(W_{opt})\,\ge
\sum_{\mu \nu }\sum_i \sum_{j \neq i}|{<w_{\mu} |\tilde{\rho}_{ij}
|w_{\mu}>}|{\cdot} |<w_{\nu} |\tilde{\rho}_{ij} |w_{\nu}>| =
\sum_i \sum_{j \neq i}\left(\sum_{\mu}|<w_{\mu}| \tilde{\rho}_{ij}
|w_{\mu}>|\right)^2\;.\label{stept} \ee Obviously: \be
\sum_{\mu}|<w_{\mu}| \tilde{\rho}_{ij}
|w_{\mu}>|\,\ge\,|\sum_{\mu}<w_{\mu}| \tilde{\rho}_{ij}
|w_{\mu}>|\;.\label{mod} \ee Then, Eq.(\ref{stept}) becomes: \be
\frac{n-1}{n}\;V^2(W_{opt})\,\ge\sum_i \sum_{j \neq
i}\left|\sum_{\mu}<w_{\mu}| \tilde{\rho}_{ij} |w_{\mu}>
\right|^2=\sum_i \sum_{j \neq i}|Tr_D
(\tilde{\rho}_{ij})|^2=\frac{n-1}{n}\,V^2\;, \label{vwv} \ee
Clearly, $V^2(W_{opt})$ becomes equal to $V$, if and only if all
the inequalities involved in the derivation of Eq.(\ref{vwv})
become equalities. Notice that the case $n = 2$ is special, for
then the Cauchy-Schwarz inequalities  Eq.(\ref{cs}) are
necessarily equalities, because the sums in Eq.(\ref{cs}) contain
just one term. However, for $n>2$, we have the equal sign  if an
only if there exist positive constants $c_{\mu}$ such that: \be
c_{\mu} \;|<w_{\mu} |\tilde{\rho}_{ij} |w_{\mu}>|= c_{\nu}
|<w_{\nu} |\tilde{\rho}_{ij} |w_{\nu}>|\;,\;\;\;\;\forall \;i \neq
j\;. \ee Since $<w_{\mu} |\tilde{\rho}_{ij} |w_{\mu}>=<w_{\mu}
|\chi_i><\chi_j|w_{\mu}>\rho_{ij}$, and we assume $\rho_{ij} \neq
0$, the above condition is equivalent to \be c_{\mu} \;|<w_{\mu}
|\chi_i><\chi_j|w_{\mu}>| = c_{\nu} \;|<w_{\nu}
|\chi_i><\chi_j|w_{\nu}>| \;,\;\;\;\;\forall \;i \neq
j\;.\label{condo} \ee On the other side, the set of inequalities
Eq.(\ref{mod}) become equalities if and only, for all $j \neq i$,
the phases of the complex numbers $<w_{\mu} |\tilde{\rho}_{ij}
|w_{\mu}>$, and then of the numbers $<w_{\mu}
|\chi_i><\chi_j|w_{\mu}>$, do not depend on $\mu$: \be
\arg\,(<w_{\mu} |\chi_i><\chi_j|w_{\mu}>)=
\theta_{ij}\;.\label{arg} \ee

\noindent  Now, for $n>2$ and $V>0$, Eq.(\ref{condo})  implies
that the matrix elements $<w_{\mu} |\chi_i>$ are all different
from zero. To see it, we separate the states $|\chi_i>$ into two
subsets, $A$ and $B$. $A$ contains the detector states which are
orthogonal to some of the eigenstates $|w_{\mu}>$. $B$ contains
the remaining states. We can prove that, for $V>0$, $A$ must be
empty. This is done in two steps: first we prove that if $A$
contains some detector states, then it contains all of them. In
the second step, we show that the elements of $A$ are orthogonal
to each other. By combining the two facts, it follows that $A$
must be empty, because otherwise all detector states would be
orthogonal to each other, and then, by taking a $W$ that has the
detector states as eigenvectors, we would achieve $D=1$ and
$V(W_{\rm opt})=0$, which is not possible, because we assumed that
$V>0$. So, let us show first that if $A$ contains some detector
states, it contains all. In fact, let $|\chi_1>$ be one of its
elements. Then there exists a value of $\mu$, say $\mu=2$, such
that $<w_2 |\chi_1>=0$. On the other side, since the vectors
$|w_{\mu}>$ form a basis for the vectors $|\chi_i>$, there must be
some eigenvector, say $|w_1>$, such that $<w_1|\chi_1> \neq 0$.
Suppose now that $B$ contains an element, say $|\chi_n>$, and
consider Eq.(\ref{condo}), for $i=1$, $j=n$, $\mu=2$ and $\nu=1$:
$c_2 \;|<w_2 |\chi_1><\chi_n|w_2>| = c_n \;|<w_1
|\chi_1><\chi_n|w_1>|$. It is clear that the l.h.s. vanishes,
while the r.h.s. does not. It follows that there cannot be such a
$|\chi_n>$. Then, if $A$ contains just one detector state, it
contains all.

\noindent Now we can turn to the second step. In order to prove
that all elements of $A$ are orthogonal to each other, consider
for example Eq.(\ref{condo}) for $\mu=2$ and $i=1$: they imply
that, for any $j \neq 1$ and any $\nu$, the numbers $|<w_{\nu}
|\chi_1><\chi_j|w_{\nu}>|$ must vanish. But this implies $<w_{\nu}
|\chi_1><\chi_j|w_{\nu}>=0$. Summing over all values of $\nu$, we
obtain: \be 0=\sum_{\nu} <w_{\nu}
|\chi_1><\chi_j|w_{\nu}>=<\chi_j|\chi_1>\;. \ee So, $|\chi_1>$ is
orthogonal to all other detector states $|\chi_i>$. The same
reasoning  applies to all elements of $A$, and thus we conclude
that all detector states are orthogonal to each other.

\noindent Having proved that all matrix elements
$<\chi_i|w_{\mu}>$ are different from zero, we can now show that
 the detector's states $|\chi_i>$ are
indeed proportional to each other.  Since $n>2$, for any $i \neq
j$, we can find a $k$ distinct from both $i$ and $j$. Consider now
Eq.(\ref{condo}) for the couples $i,k$ and $j,k$, and divide the
first by the second. This is legitimate, because  all inner
products $<w_{\mu}|\chi_i>$ are different from zero. We get: \be
\frac{|<w_{\mu} |\chi_i>|}{|<w_{\mu} |\chi_j>|} =
\;\frac{|<w_{\nu}|\chi_i
>|}{|<w_{\nu}|\chi_j>|} \;,\;\;\;\;\forall \;i \neq
j\;.\label{condt} \ee This is the same as: \be \frac{|<w_{\mu}
|\chi_i>|}{|<w_{\nu} |\chi_i>|} = \;\frac{|<w_{\mu}|\chi_j
>|}{|<w_{\nu}|\chi_j>|} \;,\;\;\;\;\forall \;i \neq
j\;.\label{condth} \ee
 Since $\sum_{\mu} |<w_{\mu} |\chi_i>|^2=1$ for all $i$, it is easy to
verify that the above equations imply: \be |<w_{\mu}
|\chi_i>|=|<w_{\mu}|\chi_j >|\;. \ee  To proceed, we make use now
of Eq.(\ref{arg}). If we set $\alpha_{\mu i}= \arg{<w_{\mu}
|\chi_i>}$, Eq.(\ref{arg}) implies: \be \alpha_{\mu i}-\alpha_{\mu
j}=\theta_{ij}\;, \ee which obviously means that, for fixed $i$
and $j$ and variable $\mu$, the phases of the complex numbers
$<w_{\mu} |\chi_i>$ and $<w_{\mu} |\chi_j>$ differ by the overall
phase $\theta_{ij}$, and this implies: \be |\chi_i>=e^{i
\theta_{ij}}|\chi_j>\;. \ee Since all detector states differ by a
phase, it obviously follows that the detector provides no
information at all, and thus $D=P$.

\section{Appendix II}
\setcounter{equation}{0}

In this Appendix, we determine the rank-one POVM that maximizes
the which-way knowledge, for the three beam interferometer
considered in Sec.V. The procedure is different, depending on
whether we choose  to measure the which-way knowledge  by means of
$K$ or ${\tilde K}$. We consider first $K$, because it is the
simplest case. We can prove then that, for any number of beams
with equal populations $\zeta_i$, and any choice of the detector
states $|\chi_i>$ in ${\cal H}_2$, the   POVM $A$ that maximizes
$K$ can be taken to have only two non vanishing elements,
$A=\{A_1,A_2\}$. The proof is as follows. First, we notice that,
for any rank-one POVM consisting of only two elements, the
conditions for a POVM, Eq.(\ref{cond}), imply: \be
\alpha_1=\alpha_2=\frac{1}{2}\;,\;\;\;\;\hat{m}_1+ \hat{m}_2=0\;.
\label{twoel} \ee Thus, all rank-one POVM with two elements are
characterized by a pair of unit vectors $\hat{m}_{\mu}$, that are
opposite to each other. Such a POVM clearly coincides with the
Projector Valued Measure (PVM) associated with the hermitian
operator $\hat{m}_1 {\cdot} \vec{\sigma}$ in ${\cal H}_2$. We let
$A$ the optimal PVM, that can be obtained by considering all
possible directions for $\hat{m}_1$.   We can show that such an
$A$ represents the optimal POVM. To see this, we prove that  the
which way-knowledge $K(A)$ delivered by $A$ is not less than that
delivered by any other POVM $C$. By virtue of the theorem proved
in Ref.\cite{davies},  it is sufficient to consider POVM's $C$
made of rank-one operators. In order to evaluate $K(C)$, it is
convenient to rewrite the quantities $p_{\mu}K_{\mu}$, for any
element $C_{\mu}=2 \alpha_{\mu}^{(C)}(1+\hat{m}_{\mu}^{(C)}\cdot
\vec{\sigma})$ of $C$, as $$
p_{\mu}K_{\mu}=\left[\frac{n}{n-1}\left(-\frac{p_{\mu}^2}{n}+\sum_{i=1}^n
\zeta_i^2 P_{i \mu}^2 \right) \right]^{1/2}=$$ \be
=\alpha_{\mu}^{(C)}\sqrt{\frac{n}{n-1}}
\left\{-\frac{1}{n}[1+(\hat{m}_{\mu}^{(C)}{\cdot}\sum_i\zeta_i
\hat{n}_i)^2]+ \sum_i \zeta_i^2[1+(\hat{m}_{\mu}^{(C)}{\cdot}
\hat{n}_i)^2]+ 2\hat{m}_{\mu}^{(C)}{\cdot}\sum_i
\zeta_i\left(\zeta_i -\frac{1}{n} \right)\hat{n}_i
\right\}^{1/2}\;. \ee We observe now that, for equally populated
beams, $\zeta_i=1/n$, the last sum in the above equation vanishes,
and the expression for $p_{\mu}K_{\mu}$ becomes invariant under
the exchange of $\hat{m}_{\mu}^{(C)}$ with $-\hat{m}_{\mu}^{(C)}$.
Consider now the POVM  $B$, such that: \be B_{\mu}^{+}
=\frac{1}{2}\;C_{\mu}\;,\;\;\;\;B_{\mu}^{-}
=\frac{1}{2}\;\alpha_{\mu}^{(C)}(1-{\hat m}_{\mu}^{(C)}{\cdot}\,
\vec{\sigma}) \ee Of course,
$p_{\mu}^{(+)}K_{\mu}^{(+)}=p_{\mu}K_{\mu}/2$, while the
invariance of $p_{\mu}K_{\mu}$ implies
$p_{\mu}^{(-)}K_{\mu}^{(-)}=p_{\mu}^{(+)}K_{\mu}^{(+)}$. It
follows that the average information for $B$ and $C$ are equal to
each other, $K(B)=K(C)$. Now, for each value of $\mu$, the pair of
operators $B^{{\pm}}_{\mu}/\alpha_{\mu}^{(C)}=(1{\pm}{\hat
m}_{\mu}^{(C)}{\cdot}\vec{\sigma})/2$ constitutes by itself a
POVM, with two elements. Thus, the POVM $C$ can be regarded as a
collection of POVM's with two elements, each taken with a
non-negative weight $\alpha_{\mu}^{(C)}$. But then $ K(C)$, being
equal to the average of the amounts of information provided by a
number of POVM with two elements, cannot be  larger than the
amount of information $K(A)$ delivered by the best POVM with two
elements. Thus we have shown that $K(C)=K(B)
\le K(A)$, which shows that   $A$ is the optimal POVM.\\
It  remains to find $A$ for the example considered in Sec.V, but
this is easy. If we let $\beta$ and $\gamma$ the polar angles that
identify the vector ${\hat m}_1$, one finds for the square of the
which-way information the following expression: \be
K^2=\frac{4}{9}  \left[ \cos^2 \beta \,\sin^2\left(
\frac{\theta}{2}\right)+3 \sin^2 \beta \cos^2 \gamma\,\cos^2
\left(\frac{\theta}{2}\right) \right]\;\sin^2\left(
\frac{\theta}{2}\right)\;. \ee For all values of $\theta$, the
which-way information is maximum if $\cos \gamma={\pm} 1$, i.e. if
the vector $\hat{m}_1$ lies in the same plane as the vectors
$\hat{n}_i$. As for the optimal value of $\beta$, it depends on
$\theta$. For $0 \le \theta < 2 \pi/3$, the best choice is $
\beta={\pm}\pi/2$, and one gets the PVM in Eq.(\ref{pvmsm}), with
gives the path distinguishability $D$ given in Eq.(\ref{smal}).
For larger values of $\theta$, one has $\beta=0$ and then the
optimal PVM is that of Eq.(\ref{opt}), with $D$ given by
Eq.(\ref{lar}).

We turn now to the case when the which-way information is measured
by means of $\tilde K$. Since the square of the predictability  is
a convex function, we are ensured by the general theorem proved in
\cite{davies} that  the optimal POVM is made of rank-one
operators, of the form (\ref{ranuno}).  We split the computation
of the optimal POVM  in two steps. First, we prove a lemma, which
actually holds for any measure of the which-way information $F$,
which is a weighted average of a convex function of the
a-posteriori probabilities $Q_{i\mu}$.
\\
\noindent {\it Lemma}: consider an interferometer with $n$ beams,
and arbitrary populations $\zeta_i$. Let the detector states
$|\chi_i>$ be in ${\cal H}_2$, and have coplanar vectors
$\hat{n}_i$. Then, the optimal POVM is necessarily such that all
the vectors ${\hat m}_{\mu}$  in Eq.(\ref{ranuno}) lie in the same
plane containing the vectors $\hat{n}_i$.

\noindent The proof of the lemma is as follows. Let $B$ be an
optimal POVM. Suppose that some of the vectors
$\hat{m}_{\mu}^{(B)}$ do not belong to the plane containing the
vectors $\hat{n}_i$, which we assume to be the $xz$ plane. We show
below how to construct a new POVM $A$ providing not less
information than $B$, and such that the vectors
$\hat{m}_{\mu}^{(A)}$ all belong to the $xz$ plane. The first step
in the construction of $A$ consists in  symmetrizing $B$ with
respect to the $xz$ plane. The symmetrization is done by replacing
each element $B_{\mu}$ of $B$, not lying in the $xz$ plane, by the
pair $(B_{\mu}^{\prime},B_{\mu}^{\prime\prime})$, where
$B_{\mu}^{\prime}=B_{\mu}/2$, and $B_{\mu}^{\prime\prime}$ has the
same weight $\alpha_{\mu}$ as $B_{\mu}^{\prime}$, while its vector
$\hat{m}_{\mu}^{(B)\prime \prime}$ is the symmetric of
$\hat{m}_{\mu}^{(B)}$ with respect to the $xz$ plane. It is easy
to verify that the symmetrization preserves the conditions for a
POVM [Eqs. (\ref{cond})]. Since all the vectors $\hat{n}_i$ belong
by assumption to the $xz$ plane, the which way knowledge actually
depends only on the projections of the vectors
$\hat{m}_{\mu}^{(B)}$ in the plane $xz$. This implies, at is easy
to check, that  symmetrization with respect to the $xz$ plane does
not change the amount of which way knowledge. We assume therefore
that $B$ has been preliminarily symmetrized in this way. Now we
show that we can replace, one after the other, each pair of
symmetric elements $(B_{\mu}^{\prime},B_{\mu}^{\prime\prime})$ by
another pair of operators, whose vectors lie in the $xz$ plane,
without reducing the information provided by the POVM. Consider
for example the pair $(B^{\prime}_{\kappa},
B^{\prime\prime}_{\kappa})$. We construct the unique pair of unit
vectors ${\hat u}_{\kappa}$ and ${\hat v}_{\kappa}$, lying the
$xz$ plane, and such that: \be {\hat u}_{\kappa}+{\hat
v}_{\kappa}=2(m_{\kappa}^{(B)x}\;\hat i+ m_{\kappa}^{(B)z}\;\hat
k)\;,\label{split} \ee where $\hat i$ and $\hat j$ are the
directions of the $x$ and $z$ axis, respectively. Notice that
${\hat u}_{\kappa}\neq{\hat v}_{\kappa}$. Consider now the
collection of operators obtained by replacing the pair
$(B^{\prime}_{\kappa},B^{\prime\prime}_{\kappa})$ with the pair
$(A_{\kappa}^{\prime},A_{\kappa}^{\prime\prime})$ such that: \be
A_{\kappa}^{\prime}=\alpha_{\kappa}^{(B)}(1+{\hat
u}_{\kappa}{\cdot} \,\vec{\sigma})\;\;\;,\;\;\;
A_{\kappa}^{\prime\prime} =\alpha_{\kappa}^{(B)}(1+{\hat
v}_{\kappa}{\cdot} \,\vec{\sigma}) \;\;. \ee It is clear, in view
of Eqs. (\ref{split}), that the new collection of operators still
forms a resolution of the identity, and thus represents a POVM.
Equations (\ref{split}) also imply:
$$
P_{i \kappa}^{(B)\prime}=P_{i \kappa}^{(B)\prime\prime}=
\alpha_{\kappa}(1+m^{(B)x}_{\kappa}n^x_i+ m^{(B)z}_{\kappa}n^z_i)=
$$
\be
=\frac{1}{2}\alpha_{\kappa}(1+u^x_{\kappa}n^x_i+
u^z_{\kappa}n^z_i)+\frac{1}{2}\alpha_{\kappa}(1+v^x_{\kappa}n^x_i+
v^z_{\kappa}n^z_i)=\frac{1}{2}(
P_{i\kappa}^{(A)\prime}+P_{i\kappa}^{(A)\prime\prime})\;,
\ee
Now, define
$\lambda^{\prime}_{\kappa}:=p^{(A)\prime}_{\kappa}/(2p_{\kappa}^{(B)})$,
 and
$\lambda^{\prime\prime}_{\kappa}:=p^{(A)\prime\prime}_{\kappa}/(2p_{\kappa}^{(B)})$,
where $p_{\kappa}^{(B)}:=
p_{\kappa}^{(B)\prime}=p_{\kappa}^{(B)\prime\prime}$. Since $
p_{\kappa}^{(A)\prime}+p_{\kappa}^{(A)\prime\prime}=2p_{\kappa}^{(B)}$, we
have $\lambda^{\prime}_{\kappa}+\lambda^{\prime\prime}_{\kappa}=1$. It is
easy to verify that:
\be
 Q_{i \kappa}^{(B)\prime}= Q_{i \kappa}^{(B)\prime\prime}=\lambda^{\prime}_{\kappa}
 \;Q_{i
\kappa}^{(A)\prime}+\lambda^{\prime\prime}_{\kappa}\;Q_{i
\kappa}^{(A)\prime\prime}\;\;,
\ee
But then, the convexity of $F$ implies:
$$
p_{\kappa}^{(B)\prime}F (\vec{Q}_{\kappa}^{(B)\prime})+
p_{\kappa}^{(B)\prime\prime}F (\vec{Q}_{\kappa}^{(B)\prime\prime})= 2
p_{\kappa}^{(B)}F (\vec{Q}_{\kappa}^{(B)\prime})=
$$
$$
=2
p_{\kappa}^{(B)}F(\lambda^{\prime}_{\kappa}\vec{Q}_{
\kappa}^{(A)\prime}+\lambda^{\prime\prime}_{\kappa}\vec{Q}_{
\kappa}^{(A)\prime\prime})\le2 p_{\kappa}^{(B)}[
\lambda^{\prime}_{\kappa}F( \vec{Q}_{ \kappa}^{(A)\prime})+
\lambda^{\prime\prime}_{\kappa}F( \vec{Q}_{ \kappa}^{(A)\prime\prime})]=
$$
\be = p_{\kappa}^{\prime(A)}F (\vec{Q}_{\kappa}^{\prime(A)})+
p_{\kappa}^{(A)\prime\prime}F
(\vec{Q}_{\kappa}^{(A)\prime\prime})\;. \ee It follows that the
new POVM is no worse than $B$. By repeating this construction, we
can obviously eliminate from $B$ all the $p$ pairs of elements not
lying in the $xz$ plane,  until we get a POVM $A$, which provides
not less information than $B$, whose elements all lie in the $xz$
plane. This concludes the proof of the lemma.\\ Now we can proceed
as follows: we consider the POVM's consisting of two elements
only, and having its vectors ${\hat m}_i$ parallel to the $x$
axis.  By direct evaluation one can check that $\tilde{K}(A)$
equals the expression in Eq.(\ref{smal}). We can prove that, for
$0 \le \theta < 2 \pi/3$, such an $A$ provides not less
information than any other POVM, $C$, consisting of more than two
elements. By virtue of the lemma just proven, we loose no
generality if we assume that the  all the vectors $m_{\mu}^{(C)}$
of $C$ lie in the $xz$ plane. Our first move is to symmetrize $C$
with respect to $z$ axis, by introducing a POVM $B$, consisting of
pairs of elements $(B_{\mu}^{\prime}, B_{\mu}^{\prime\prime})$,
having equal weights, and vectors $\hat {m}_{\mu}^{\prime}$ and
$\hat {m}_{\mu}^{\prime\prime}$ that are symmetric with respect to
the $z$ axis: \be
B^{\prime}_{\mu}=\frac{1}{2}\;C_{\mu}\;\;,\;\;\;\;B^{\prime\prime}_{\mu}=
\frac{1}{2}\;\alpha_{\mu}^{(C)}(1-{\hat m}_{\mu}^x{\sigma}_x+
{\hat m}_{\mu}^z{\sigma}_z)\;\;\;,\;\;\; \ee $B$ provides as much
information as $C$. Indeed, in view of Eq. (\ref{easy}), we find
\be P_{{\pm} \mu}^{(C)}=2\;P_{{\pm} \mu}^{(B)\prime}=2\;P_{\mp
\mu}^{(B)\prime\prime}\;\;\;,\;\;\; \ee The invariance of the
predictability with respect to permutations of its arguments, then
ensures that $\tilde{K}(B)=\tilde{K}(C)$. Thus, we loose no
information if we consider a POVM $B$, that is symmetric with
respect to the $z$ axis. Now we describe a procedure of reduction
that, applied to a symmetric POVM like $B$, gives rise to another
symmetric POVM $\hat B$, which contains two elements less than
$B$, but nevertheless  gives no less information than $B$. The
procedure works as follows: we pick at will two pairs of elements
of $B$, say $(B_{N}^{\prime}, B_{N}^{\prime\prime})$ and
$(B_{N-1}^{\prime}, B_{N-1}^{\prime\prime})$ and consider the
unique pair of symmetric unit vectors $\hat{u}_{{\pm}}={\pm}u^x
\;\hat i + u^z\; \hat k$ such that: \be
{u}^z=\frac{1}{\alpha_N^{(B)}+\alpha_{N-1}^{(B)}}
(\alpha_N^{(B)}\; m_{N}^{(B) z} +
 \alpha_{N-1}^{(B)}\; m_{N-1}^{(B)z})\;.\label{vecu}
\ee
Consider the symmetric collection $\hat B$, obtained from $B$ after
replacing the four elements $(B_{N}^{\prime},
B_{N}^{\prime\prime},B_{N-1}^{\prime}, B_{N-1}^{\prime\prime})$ by the pair
$(\hat{B}_{N-1}^{\prime},\hat{B}_{N-1}^{\prime\prime})$ such that:
\be
\hat{B}_{N-1}^{\prime}=(\alpha_N^{(B)}+\alpha_{N-1}^{(B)})(1+
\hat{u}_+ {\cdot}\, \vec{\sigma})\;,\;\;\;\;
\hat{B}_{N-1}^{\prime\prime}=(\alpha_N^{(B)}+\alpha_{N-1}^{(B)})(1+
\hat{u}_- {\cdot} \,\vec{\sigma})\;.
\ee
$\hat B$ is still a POVM, as it is easy to verify. Moreover, $\hat B$
provides not less information than $B$, as we now show. Indeed, after some
algebra, one finds:
\be
\frac{\tilde{K}(\hat B)-\tilde{K}(B)}{\alpha_N^{(B)}+\alpha_{N-1}^{(B)}}=
 g(u^z)-
\frac{\alpha_N^{(B)}}{\alpha_N^{(B)}+\alpha_{N-1}^{(B)}}g(m_N^{(B)z})-
\frac{\alpha_{N-1}^{(B)}}{\alpha_N^{(B)}+\alpha_{N-1}^{(B)}}
g(m_{N-1}^{(B)z}) \;,\label{last} \ee where the function $g(x)$
has the expression: \be g(x)=-\frac{3+x(1+2 \cos
\theta)}{6}+\frac{(1+x)^2+2(1+x \cos \theta)^2+2(1-x^2)\sin^2
\theta}{6+2x(1+2\cos \theta)} \;. \ee In view of Eq. (\ref{vecu}),
the r.h.s. of Eq. (\ref{last}) is of the form \be g(\lambda x_1 +
(1-\lambda)x_2)-\lambda \,g(x_1) -(1-\lambda)
\,g(x_2)\;,\label{form} \ee where
$\lambda=\alpha_N^{(B)}/(\alpha_N^{(B)}+\alpha_{N-1}^{(B)})$,
while $x_1=m_N^{(B)z}$ and $x_2=m_{N-1}^{(B)z}$. It may be checked
that, for all values of $\theta$, such that $0 \le \theta <\,2
\pi/3$, $g(x)$ is concave, for $x\in [-1,1]$, and so the r.h.s. of
Eq. (\ref{form}) is non-negative for any value of $\lambda \in
[0,1]$. This implies that the r.h.s. of Eq. (\ref{last}) is
non-negative as well,
 and so $\tilde{K}(\hat B)\ge
\tilde{K}(B)$. After  enough iterations of this procedure, we end up with a
symmetric POVM consisting of two pairs of elements $(B_{1}^{\prime},
B_{1}^{\prime\prime})$ and $(B_{2}^{\prime}, B_{2}^{\prime\prime})$. But
then, the conditions for a POVM, Eqs. (\ref{cond}), imply that the quantity
between the brackets on the r.h.s. of Eq. (\ref{vecu}) vanishes,  and so
Eq. (\ref{vecu}) gives $u^z=0$. This means that the last iteration gives
rise precisely to the PVM $A$.   By  putting everything together, we have
shown that $\tilde{K}(C) = \tilde{K}(B)\le
\tilde{K}(\hat{B})\dots \le \tilde{K}(A)$, and this is the required result.

\end{document}